\documentclass{epl}
\usepackage{latexsym,epsfig}

\newcommand{\ve}{\vert}
\newcommand{\fr}{\frac}
\newcommand{\beq}{\begin{equation}}
\newcommand{\eeq}{\end{equation}}

\title{Localized low-frequency Neumann modes in 2d-systems with rough boundaries}
\author{S. Russ\inst{1} and Y. Hlushchuk\inst{1}}
\institute{
  \inst{1} Institut f\"ur Theoretische Physik III, Universit\"at Giessen,
D-35392 Giessen, Germany
}


\begin{document}

\maketitle

\begin{abstract} 
We compute the relative localization volumes of the vibrational eigenmodes 
in two-dimensional systems with a regular body but 
irregular boundaries under Dirichlet and under Neumann 
boundary conditions. We find that localized states are rare under 
Dirichlet boundary conditions but very common in the Neumann case. In order 
to explain this difference, we utilize the fact that under Neumann conditions the 
integral of the amplitudes, carried out over the whole system area is zero. We 
discuss, how this condition leads to many localized states in the low-frequency regime 
and show by numerical simulations, how the number of the localized states 
and their localization volumes vary with the boundary roughness.
\end{abstract}

\section{Introduction}

The problem of localization in disordered systems \cite{kramer} has been subject 
to intense research for several decades.
Localization of vibrational or electronical eigenmodes 
has turned out to change the physical properties, 
as e.g. the electric properties \cite{kramer,elect,anderson},
the damping of acoustic resonators \cite{damping} and cavities 
\cite{acoust}, the band gap in semiconductors \cite{poroes}, the wave transport
\cite{transport} and the density of vibrational states \cite{DOS1,DOS2}.
The reasons for localization have been discussed for long and 
it is generally accepted that localization always involves some 
geometrical or structural irregularity of the system. 
This irregularity can be present as bulk irregularity or as boundary roughness. 
The first case has been widely
investigated and standard models are e.g. the Anderson model 
\cite{kramer,anderson} and the percolation model \cite{perk1}. Less works 
however addressed the problem of
localization in systems with an ordered bulk material and an irregular boundary. 
One model for this problem is the model of fractal 
drums, where localized states have been found under Dirichlet \cite{Even,DOS2} 
as well as under Neumann \cite{DOS1,DOS2} boundary conditions.

Recently \cite{HlRu2003}, localized states have also been found in 
$2d$-systems with non-fractal boundary roughness under Neumann boundary 
conditions, while under Dirichlet conditions they are rare and seem to be 
linked to the existence of confined regions (which indeed occur in the case 
of fractal drums \cite{Even} or in systems with hard scatterers \cite{sridhar}).
In this Letter we focus on this phenomenon and
consider several types of systems with non-fractal irregular surfaces. 
Under Neumann boundaries we find 
many localized states in all systems with arbitrarily shaped irregular
boundaries. We elucidate this behavior by numerical simulations and show 
how it can be explained by using a sum-rule for the Neumann case.

\section{Localized Modes}

We start with the Helmholtz equation for the vibrational amplitude 
$\psi_\alpha(x,y)$ of a membrane that is located
in the $xy$-plane and vibrates in $z$-direction,
\begin{equation} \label{helmholtz}       
\Delta \psi_\alpha(x,y) = -\fr{\omega_\alpha^2}{c^2} \psi_\alpha(x,y),       
\end{equation}   
with the state index $\alpha$, the eigenfrequency $\omega_\alpha$ and the 
sound velocity $c$. The boundary may have
an arbitrary shape and may either be fixed (Dirichlet case) or 
vibrate freely (Neumann case). The latter case applies e.g. for gas vibrations in
acoustic resonators, where $\psi$ describes the pressure 
distribution of the gas \cite{acoust}.
For calculating the eigenfunctions and -frequencies, we
discretize Eq.~(\ref{helmholtz}) on a square lattice, where nearest-neighbor masses 
$m$ are coupled by linear springs $k$ and diagonalize it by the
Lanczos algorithm. 
When the number of masses tends to infinity, we approach 
a continuous system, where $a^2k/m\to c^2$, with the lattice constant $a$.

\unitlength 1.85mm
\vspace*{0mm}
{
\begin{figure}
\begin{picture}(80,38)
\def\epsfsize#1#2{0.40#1}
\put(56,20){\epsfbox{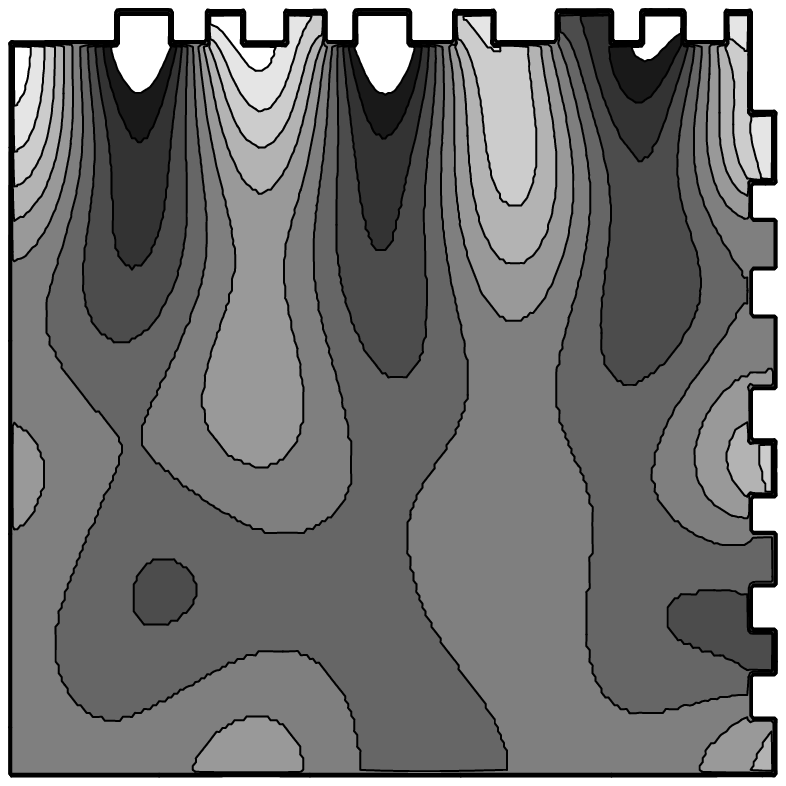}}
\put(56,0){\epsfbox{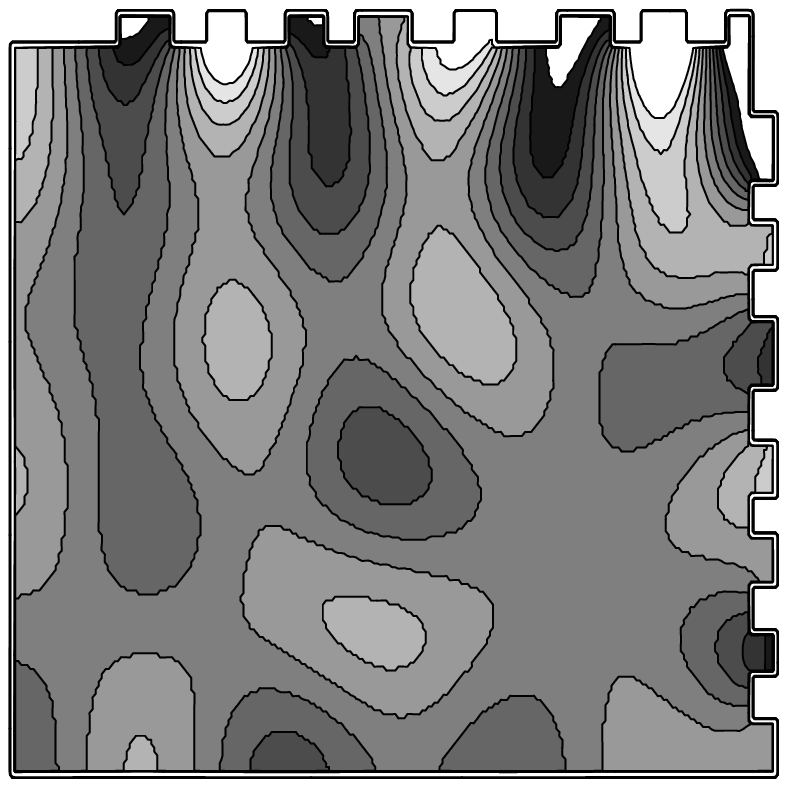}}
\def\epsfsize#1#2{0.45#1}
\put(28,20){\epsfbox{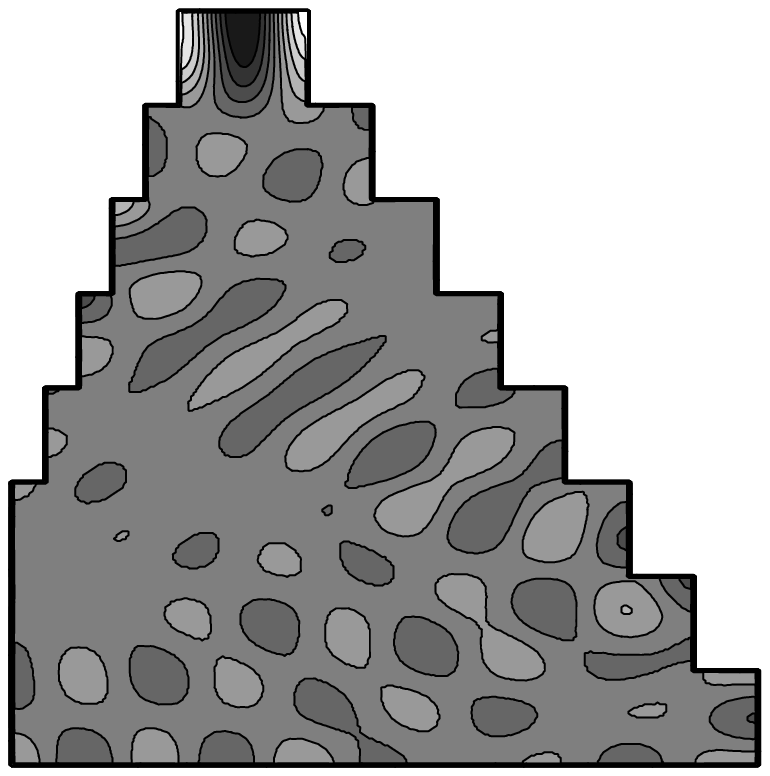}}
\put(28,0){\epsfbox{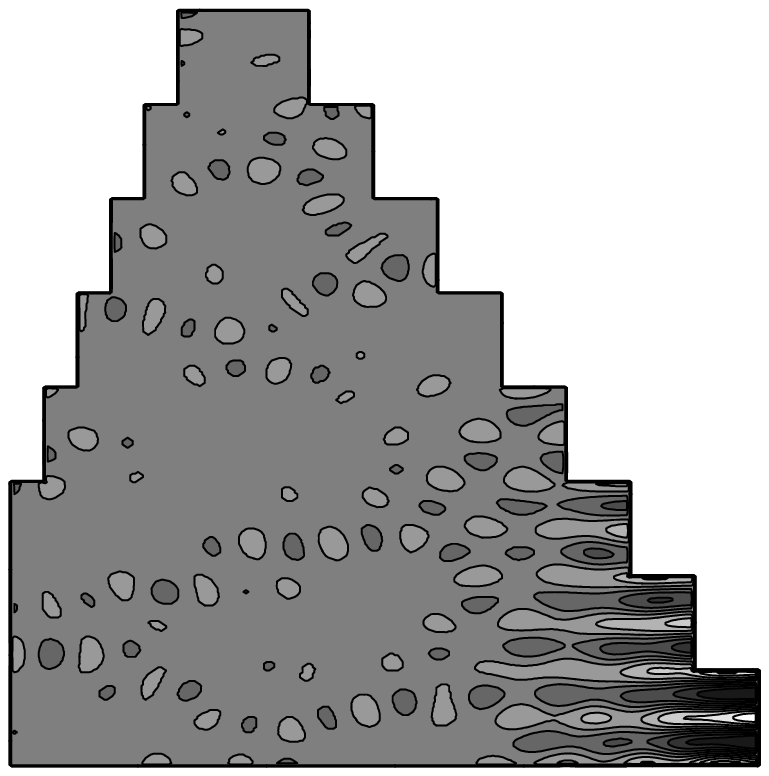}}
\put(2,0){\epsfbox{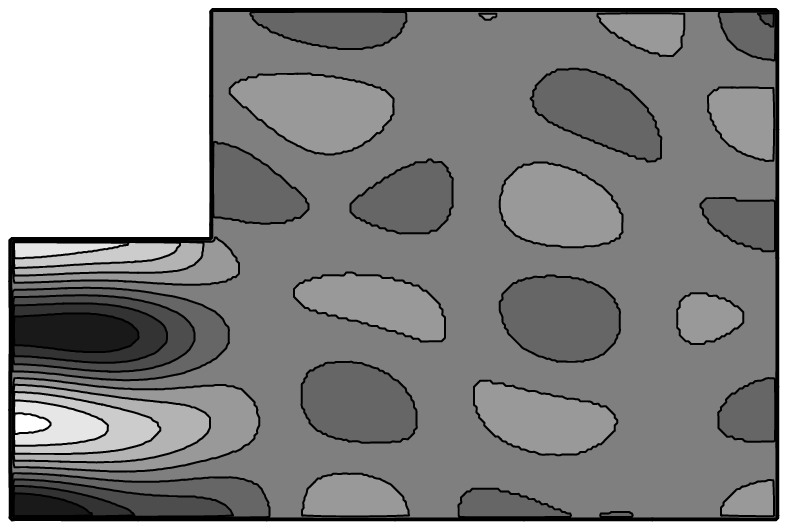}}
\put(2,20){\epsfbox{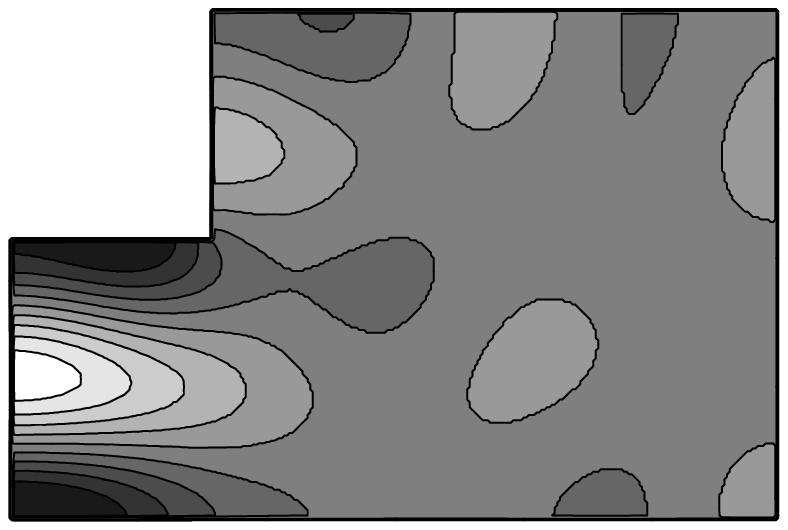}}
\put(4,38){\makebox(1,1){\bf\Large (a)}} 
\put(27,38){\makebox(1,1){\bf\Large (c)}}  
\put(52,38){\makebox(1,1){\bf\Large (e)}}                    
\put(4,15){\makebox(1,1){\bf\Large (b)}}
\put(27,15){\makebox(1,1){\bf\Large (d)}} 
\put(52,15){\makebox(1,1){\bf\Large (f)}}                     
\end{picture}
\caption[]{\small Localized modes of several systems with rough boundaries
under Neumann boundary conditions for different system geometries, (a-b)
L-shaped system, (c-d) triangular systems and (e-f) tooth systems.
The amplitudes are indicated by different gray levels.
The white and black regions stand for positive and negative amplitudes, 
respectively and the neutral gray tone stands for nearly zero amplitude. 
The eigenfrequencies and relative localization volumes are 
(a) $\omega\approx 0.060\,(k/m)^{1/2}$, $V_{\rm{loc}}\approx 0.13$, 
(b) $\omega\approx 0.088\,(k/m)^{1/2}$, $V_{\rm{loc}}\approx 0.13$, 
(c) $\omega\approx 0.125\,(k/m)^{1/2}$, $V_{\rm{loc}}\approx 0.07$, 
(d) $\omega\approx 0.226\,(k/m)^{1/2}$, $V_{\rm{loc}}\approx 0.06$, 
(e) $\omega\approx 0.072\,(k/m)^{1/2}$, $V_{\rm{loc}}\approx 0.17$ and 
(f) $\omega^2\approx 0.083\,(k/m)^{1/2}$, $V_{\rm{loc}}\approx 0.09$.
}
\label{bi:neu1}
\end{figure}}

For regular systems, all states are extended for both types of 
boundary conditions. The
eigenstates of a rectangle of side lengths $L_x$ and $L_y$, e.g.,
are simple sine- or cosine functions under Dirichlet and under
Neumann boundary conditions, respectively. 
For a discretized rectangle of $Z_x=L_x/a$ and 
$Z_y=L_y/a$ lattice segments
along the $x$- and the $y$-axis we have 
$\omega^2_{n,m}=(2k/m)\left(2-\cos(n\pi/\ell_x) - \cos(m\pi/\ell_y)
\right)$
where $\ell_i=Z_i+1$, $n\in[0,Z_x]$, $m\in[0,Z_y]$ under Neumann and 
$\ell_i=Z_i$, $n\in[1,Z_x-1]$, $m\in[1,Z_y-1]$ under Dirichlet boundary
conditions \cite{DOS1,born}.
In the continuous limit $n\ll Z_x, m\ll Z_y$ the eigenvalues converge to 
$\omega^2_{n,m} = \pi^2c^2(n^2/L_x^2+m^2/L_y^2)$.

For systems with irregular boundaries, the eigenstates show an irregular 
pattern of several mountains and valleys of similar but not identical shapes, 
whose extensions refer roughly to half a wavelength $\lambda$. Therefore,
at higher frequencies the mountains and valleys are narrower and
therefore more numerous than at low frequencies.
This can be seen in Fig.~\ref{bi:neu1}(a-f), where some representative localized
states of different systems are shown. 
For the localized states of Fig.~1, 
the absolute values of the vibrational amplitudes are only large at one part 
of the system and decay rapidly towards the other system side. 

In order to quantify the degree of localization, we determined numerically the 
relative localization volumes $V_{\rm{loc}}^{(\alpha)}$ (participation ratio) 
\cite{wegner},
\begin{equation}
V_{\rm{loc}}^{(\alpha)}={V_\alpha \over A} \equiv \frac{1}
{A\,\int\left|\psi^{(\alpha)}\right|^4\,dx dy},
\end{equation}
where $\psi$ is normalized according to $\int\left|\psi(x,y)\right|^2dxdy=1$
and $A$ is the membrane area.
$V_{\rm{loc}}$ is equal to one for constant functions, whereas for regular sine
or cosine-functions of an ordered square membrane we have $V_{\rm{loc}}=1$ for 
$n=m=0$, $V_{\rm{loc}}=2/3$ for $n$ or $m=0$ and $V_{\rm{loc}}=4/9$ for $n,m>0$.
For a Gaussian function, we have $V_{\rm{loc}}=1/3$. If $V_{\rm{loc}}$ is
much smaller than this value, we call the function localized. 
For comparison, the values of $V_{\rm{loc}}$ are given in the caption of
Fig.~\ref{bi:neu1}.
In Fig.~\ref{bi:locvol} we show all values of $V_{\rm{loc}}$ of the
low-frequency modes of the systems of Fig.~\ref{bi:neu1} for (a) Dirichlet and
(b) Neumann boundary conditions. In Fig.~\ref{bi:locvol}(a) small values  
$V_{\rm{loc}}$ below $0.2$ do not occur and most values are above $0.33$.
In Fig.~\ref{bi:locvol}(b) on the other hand, the distribution of the 
localization volumes is much broader and shows a tail, with a large
number of values even below $0.1$.
So, there must be a strong localization mechanism in the Neumann case, which 
apparently does not exist under Dirichlet boundary conditions.

\unitlength 1.85mm
\vspace*{0mm}
{
\begin{figure}
\begin{picture}(80,25)
\def\epsfsize#1#2{0.30#1}
\put(5,0){\epsfbox{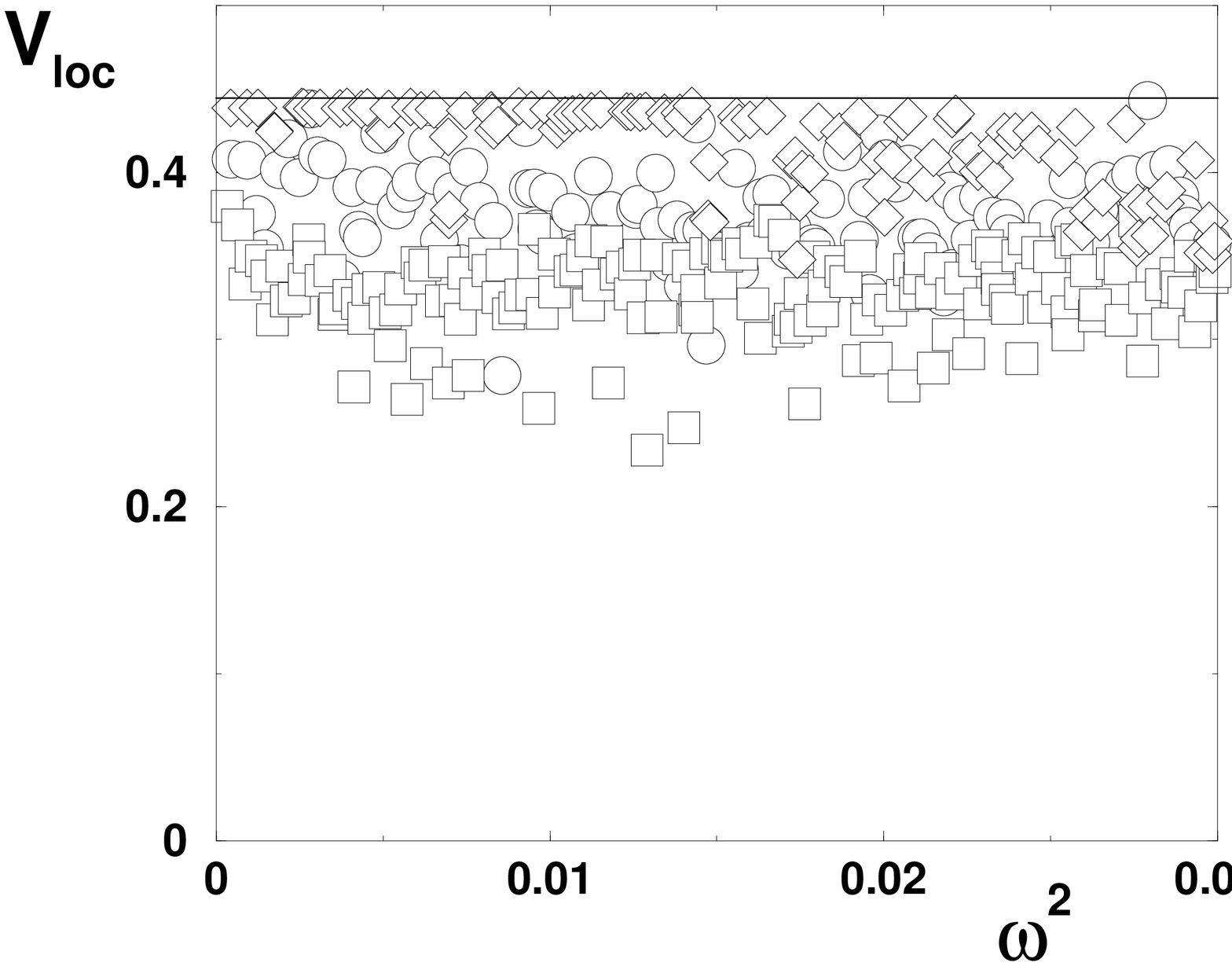}}
\put(45,0){\epsfbox{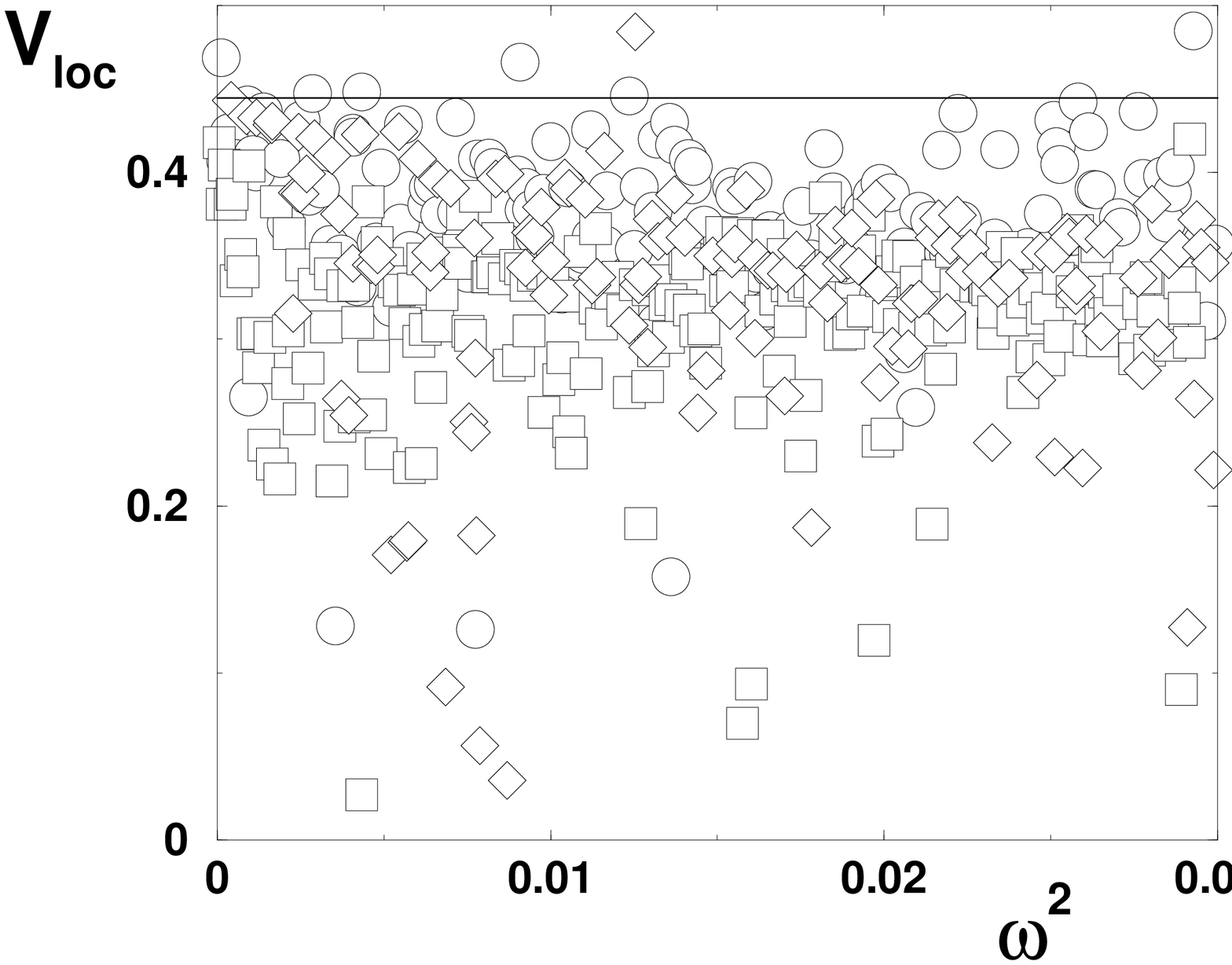}}
\put(28,6){\makebox(1,1){\bf\Large (a)}} 
\put(68,6){\makebox(1,1){\bf\Large (b)}}      
\end{picture}
\caption[]{\small Relative localization volumes $V_{\rm{loc}}$ versus $\omega^2$ for 
the systems of 
Fig.~\ref{bi:neu1} under (a) Dirichlet and (b) Neumann boundary conditions. 
The symbols $\circ$, $\Box$ and $\diamond$ stand for the geometries of
Fig.~\ref{bi:neu1}(a-b) (L-shaped system), (c-d) (triangular system)
and (e-f) (tooth system), respectively.
The straight line refers to the value of $4/9$ of
an ordered square membrane. 
}
\label{bi:locvol}
\end{figure}}

To understand this difference, we recall that a sum rule
applies in the Neumann case for each wavefunction (except for the
first one, where $\omega_\alpha=0$) \cite{DOS1}. This can be seen by
integrating Eq.~(\ref{helmholtz}) over $dx\,dy$ and applying the
divergence theorem, yielding
$-(\omega_\alpha^2/c^2) \int\int dx dy \,     
\psi_\alpha(x,y) = \int_\Gamma (\nabla\psi_\alpha) \vec n ds$,      
where $\Gamma$ is the boundary of the membrane, $\vec n$ its normal
component and the double integral   
on the left-hand side is carried out over the membrane area. 
With the Neumann condition $\partial\psi/\partial \vec n =
(\nabla\psi)\vec n=0$ along $\Gamma$, we find the sum rule
\beq\label{cond}
\int\int\,dxdy\,\psi(x,y)=0 \qquad\mbox{for}\quad\omega\ne 0, 
\eeq 
which is rather restrictive. As we will see in the following
it strongly limits the way, how a wavefunction that is well-suited to a given 
rough boundary evolves into the rest of the system and is the key in 
understanding the occurence of the localized states under
Neumann conditions. In the Dirichlet case, where Eq.~(\ref{cond}) does not apply, 
the absolute value of the sum $\int\int\,dxdy\,\psi(x,y)$ can take any value, 
i.e. the mountains and valleys of the mode do not need to cancel each other.

\unitlength 1.85mm
\vspace*{0mm}
{
\begin{figure}
\begin{picture}(80,13)
\put(0,0.5){\unitlength0.3mm
\linethickness{0.4pt} 
\put(0,0){\line(1,0){110}}
\put(0,41){\line(1,0){30}}
\put(30,79){\line(1,0){80}}
\put(0,0){\line(0,1){41}}
\put(110,0){\line(0,1){79}}
\put(30,41){\line(0,1){38}}
\multiput(30,0)(0,8){6}{\line(0,1){4}}
\put(15,12){\makebox(0,0)[b]{\huge A}}
\put(70,30){\makebox(0,0)[b]{\huge B}}
}
\def\epsfsize#1#2{0.25#1}
\put(42,0){\epsfbox{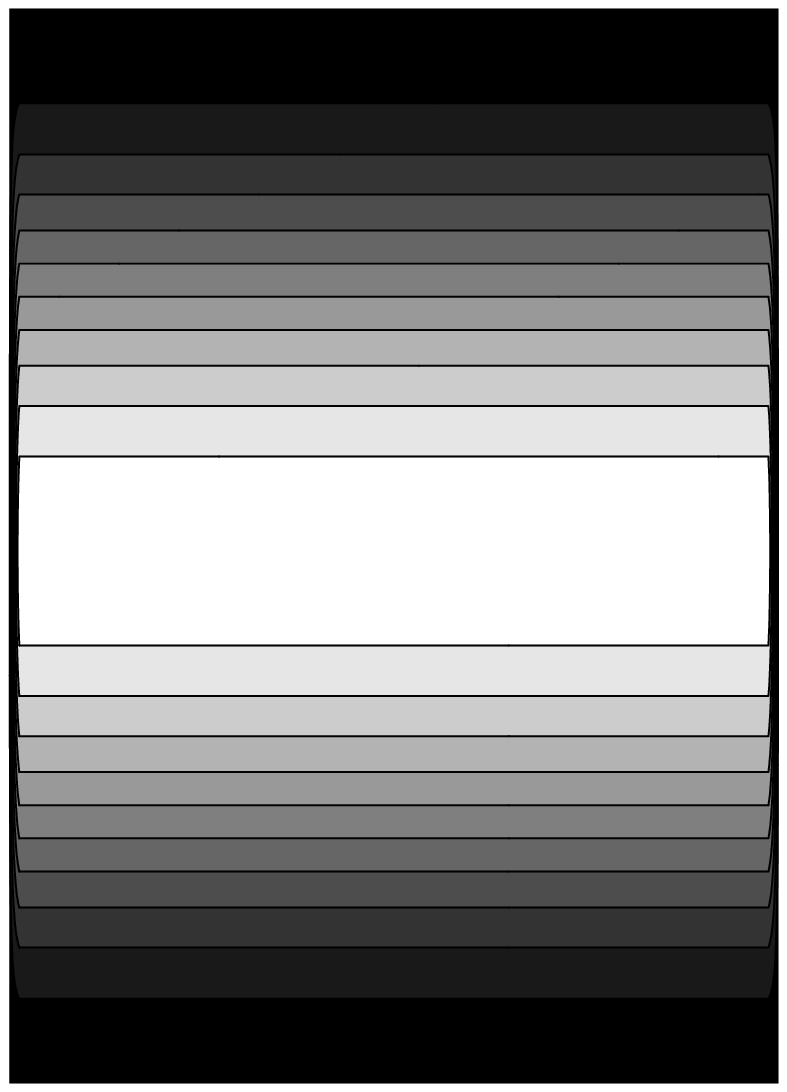}}  
\put(25,0){\epsfbox{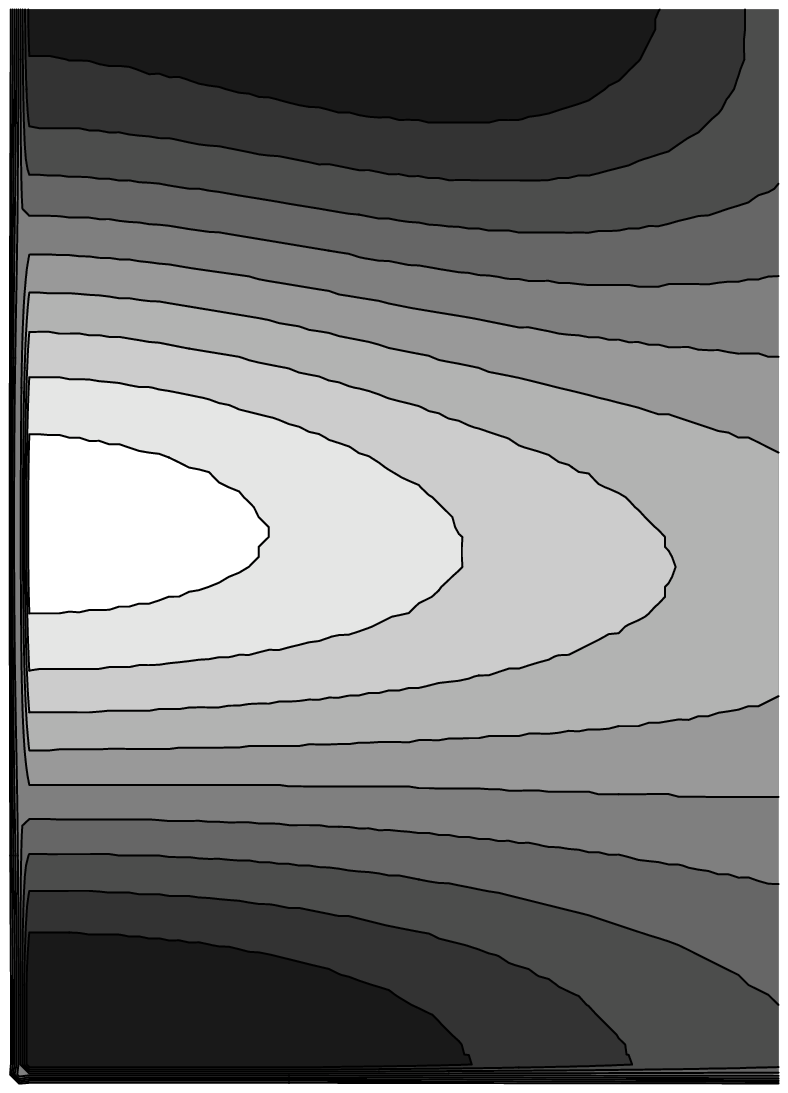}}  
\def\epsfsize#1#2{0.5#1}
\put(55,0){\epsfbox{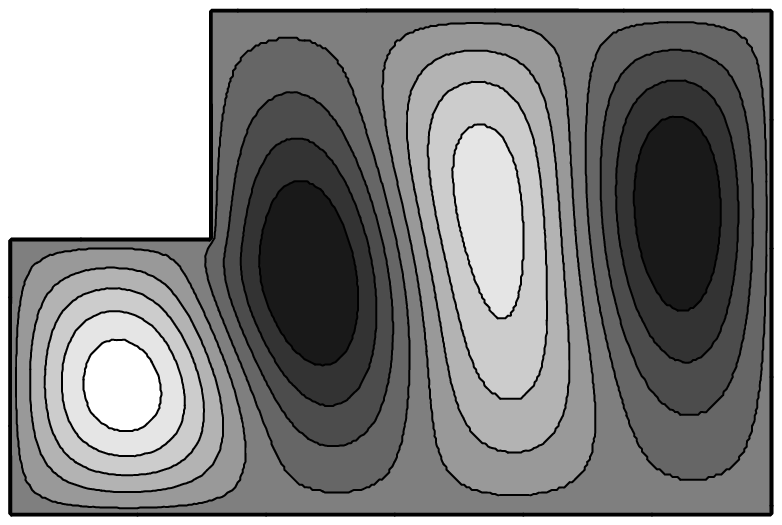}}
\put(0,12){\makebox(1,1){\bf\Large (a)}}    
\put(22,12){\makebox(1,1){\bf\Large (b)}}  
\put(39,12){\makebox(1,1){\bf\Large (c)}} 
\put(57,12){\makebox(1,1){\bf\Large (d)}}                       
\end{picture}
\caption[]{\small The L-shaped system $AB$. (a) The geometry. 
(b) 20th Eigenmode of the total system $AB$ in part $A$ alone.
(c) Eigenmode $(n,m)=(0,2)$ of the isolated rectangle $A$. 
(In (b,c), part $A$ is shown enlarged as compared to (a).) 
(d) Example of the 6th eigenmode under Dirichlet boundary conditions,
arising from the $(1,1)$-mode of rectangle $A$ for comparison. 
}
\label{bi:2step}
\end{figure}}

\section{The L-shaped system}
To show, how the sum rule~(\ref{cond}) leads to localized states, 
we first consider the L-shaped system, which is
composed of two rectangles $A$ and $B$ of sizes $A_x\times A_y$ and 
$B_x\times B_y$, respectively (see Fig.~\ref{bi:2step}(a)). We assume that 
$A_x$ and $A_y$ are no divisors of $B_x$ and $B_y$ (see below),
i.e. $B_x\ne iA_x$ and $B_y\ne jA_y$, $i,j \in N$. 
Figure~\ref{bi:2step}(b) shows the eigenstate of Fig.~\ref{bi:neu1}(a), 
in $A$ alone (zoomed to a larger size). Comparing it
to the state $(n,m)=(0,2)$ of the isolated rectangle $A$ (cf.~Fig.~\ref{bi:2step}(c)),
we see that apart from some decay from the left towards 
the right side of the system in (b), both mode patterns are quite similar.
Also the frequencies $\omega$ of both states are very close, as can be seen in
Table~1 ($(n,m)=(0,2)$, $\alpha=20$). 
Also many other rectangular eigenstates, with $n=0$, $m>0$ show up in the 
spectrum of the L-shaped system with nearly zero amplitudes in region $B$
\cite{footnote}. 

\begin{table}  
\begin{tabular}{|c|c||c|c|c|}
\hline 
\multicolumn{5}{|c|}{\rule[-2mm]{0mm}{6mm}(i)}\\
\hline 
{\rule[-3mm]{0mm}{8mm}$(n,m)$} & $\omega_{n,m}^{\rm{Neu}}$ & $\alpha^{\rm{Neu}}$& 
$\omega^{\rm{Neu}}_\alpha$ & $V_{\rm{loc},\alpha}^{\rm{Neu}}$\\
\hline
 (0,1) & 0.0286 &  7 &0.0307 & 0.265 \\
  (0,2) & 0.0571 &  20 &0.0596 & 0.128 \\
  (0,3) & 0.0857&  40 &0.0880& 0.126 \\
  (0,4) & 0.1142&  67 &0.1166& 0.158 \\
  (0,5) & 0.1427&  99 &0.1447& 0.259 \\
  (0,6) & 0.1712&  138 &0.1733& 0.142 \\
\hline
\end{tabular}
\hspace{1cm}
\begin{tabular}{|c|c||c|c|c|}
\hline 
\multicolumn{5}{|c|}{\rule[-2mm]{0mm}{6mm}(ii)}\\
\hline 
{\rule[-3mm]{0mm}{8mm}$(n,m)$} & $\omega_{n,m}^{\rm{Dir}}$ & $\alpha^{\rm{Dir}}$& 
$\omega^{\rm{Dir}}_\alpha$ & $V_{\rm{loc},\alpha}^{\rm{Dir}}$
 \\
\hline
 (1,1) & 0.0495 &  6 &0.0464 & 0.420 \\
  (1,2) & 0.0703 &  16 &0.0700 & 0.423 \\
  (1,3) & 0.0954&  29 &0.0925& 0.280 \\
  (1,4) & 0.1221&  54 &0.1248& 0.402 \\
  (1,5) & 0.1495&  82 &0.1500& 0.409 \\
  (1,6) & 0.1773& 113 &0.1766& 0.322 \\
\hline
\end{tabular}
\bigskip
\caption[]{\small Comparison of the eigenfrequencies $\omega_{n,m}$ 
(in units of $(k/m)^{1/2}$) of 
rectangle $A$ with the side lengths (numbers of particles)
$N_x=79, N_y=110$ and $\omega_\alpha$ of the 
L-shaped system for (i) Neumann and (ii) Dirichlet boundary
conditions. 
The last columns show $V_{\rm{loc},\alpha}$ of the L-shaped system.
}\label{table1}
\end{table}

We now show how the localization is related to the sum 
rule~(\ref{cond}): If modes, originating from regular modes in $A$ 
are extended, they must evolve into region $B$. 
An integer number of the mountains and valleys, that are optimized for 
region $A$ must travel to region $B$, satisfying (i) Eq.~(\ref{helmholtz}) 
at each lattice point and (ii) the sum rule (\ref{cond}).
This can be achieved as follows: The integral 
$S_A\equiv\int_A\int\,dxdy\,\psi(x,y)$ over region $A$ alone is close to
zero, as it arises from an eigenmode of rectangle $A$. In region $B$,
the corresponding integral $S_B\equiv\int_B\int\,dxdy\,\psi(x,y)$ must 
therefore just compensate the value of $S_A$,
i.e. $S_B = -S_A$. As $S_A$ is small, it is clear that this condition is 
easier to fulfill, when 
the absolute values of $\psi$ in region $B$ are small. When the 
$\ve\psi(x,y)\ve$ are large in region $B$, the single mountains and valleys 
need to cancel each other accurately, which is in most cases impossible because of
the mismatch between the two rectangles $A$ and $B$. If on the other hand the
eigenfunction decays to very small values in region $B$, also the integral 
$S_B$ stays small. Moreover, by a suited decay rate 
it can always be adjusted such that $S_B=-S_A$, thereby fulfilling the sum 
rule~(\ref{cond}). This clearly leads to localized modes.

Under Dirichlet boundary conditions, the situation is different, as the sum
rule (\ref{cond}) is not valid and therefore the possible modes are much less
restricted. 
One typical mode is shown in Fig.~\ref{bi:2step}(d), which
arises from the $(n,m)=(1,1)$-mode of rectangle $A$.
Obviously, this mode is extended and clearly, 
$\int\int\psi(x,y) dx dy$ is unequal to zero.

To test this assumption, we analyzed the mode pattern and the eigenvalues
of the modes, arising from rectangle $A$ under (i) Neumann and (ii) Dirichlet 
conditions. In Table~1, we compare their 
frequencies $\omega_\alpha$ and relative localization volumes $V_{\rm{loc},\alpha}$
to the corresponding frequencies $\omega_{n,m}$ of 
rectangle $A$. In the Neumann case, most of them are indeed 
localized. (Not
shown here, there are also localized modes of slightly higher localization 
volumes arising from the other two
possible rectangles of the upper and the lower part of the system.) 
Most Dirichlet modes, on the other hand, possess much higher values of $V_{\rm{loc}}$.

\section{Localization in general systems}

We now consider more complex systems, as e.g. the systems of 
Fig.~\ref{bi:neu1}(c-f). The localization in these cases can be understood 
by the same mechanism as above, i.e. by eigenmodes of some boundary 
subsystems whose amplitudes decrease when entering other regions 
of the system in order to fulfill the sum rule~(\ref{cond}).
This can be seen in Fig.~\ref{bi:neu1}, where we can 
most easily identify rectangular eigenfunctions in boundary
structures (c-d), or modes of several close-lying 
boundary rectangles that couple (e-f). 

We now discuss, how the number of localized states and the values of the
smallest localization volumes
can be influenced by the boundary. To this end, we calculated the histograms of
the $V_{\rm{loc}}$ of several systems under both types of boundary conditions 
in the frequency regime of $0<\omega^2<0.4\,k/m$. 
Figure~\ref{bi:hists}(a) shows as an example the complete
histogram of the tooth system under Neumann conditions, which
has a strong maximum at $V_{\rm{loc}}\approx 
0.33$, which refers to a Gaussian distribution of the amplitudes and does not 
change when going to other systems or boundary conditions.

We are mostly interested in the tails of the histograms towards 
smaller values, which depend sensibly on the boundary details. 
In Fig.~\ref{bi:hists}(b-e) we show the histogram tails for $0\le 
V_{\rm{loc}}\le 0.3$ for (b) the L-shaped system, (c) the triangular system, 
(d) the tooth system and
(e) a ''confined'' L-shaped system where region $A$ was made much longer and 
narrower. The histograms for the Neumann states 
are indicated by white columns and the histograms for Dirichlet states
by black columns. 

We can see that in the first three cases, localized states under Neumann conditions are much 
more frequent than under Dirichlet conditions. 
In the normal L-shaped system we found only two modes with $V_{\rm{loc}} <0.2$ in 
the given frequency range in the Dirichlet case, but
a large number of localized states in the Neumann case with $V_{\rm{loc}}$ going 
down to values even smaller than $0.1$. 
This phenomenon is even stronger in the case of the triangular system of
Fig.~\ref{bi:hists}(c) and most pronounced 
in Fig.~\ref{bi:hists}(d), which corresponds to the system with many narrow 
boundary ''teeth'' and where no localized states are found in the Dirichlet
case.

In the Neumann case, the number and $V_{\rm{loc}}$-values of the localized states change with the shape of the systems.
First, as complex boundaries possess more small substructures, where localized eigenfunctions can exist, the number of localized states increases with the complexity of the boundary, i.e., the systems of Figs.~\ref{bi:hists}(c) and (d) possess more localized states than the L-shaped systems. (Note the different range of the $y$-axis.)
Second, the smallest values of $V_{\rm{loc}}$ decrease with the relative size of the boundary substructures, i.e. the values of $V_{\rm{loc}}$ in Figs.~\ref{bi:hists}(c,d) decrease to smaller values than in (b,e). 
Third, also the total system area $A$ influences the relative number of localized states. 
Whereas the number of extended states grows $\propto A$, the number of substructures and thus of the localized states grows $\propto \sqrt{A}$ (given a constant density of substructures along the boundary). 
Therefore, the relative number of localized states should become smaller with increasing system size. 
Nevertheless, the number of localized states should stay non-negligiable in most cases of rough or porous materials that are built up by smaller clusters that do not grow to arbitrarily large sizes. 
These relations in mind, one could use the localization behavior also for technical applications, as e.g. acoustic damping in systems where the surface is covered with an absorbing material. It has already been shown that localized states are much more strongly damped than extended states \cite{damping,acoust}, so that a suitable choice of the size of the substructures could be very efficient in damping the modes of a certain frequency range.

Only the confined L-shape system of Fig.~\ref{bi:hists}(e) is exceptional, 
as it possesses localized states with values of $V_{\rm{loc}}$ close 
to $0.1$ also under Dirichlet conditions. 
This is in agreement with earlier numerical and experimental results on fractal
drums \cite{Even} and on systems with hard scatterers \cite{sridhar}, where 
localized states under Dirichlet conditions have also been 
found in confined regions. Clearly, localization under Dirichlet
conditions is not forbidden, but it is quite rare and the localization
mechanism is not the same as the one described here.
Whereas under Neumann conditions, slight boundary
irregularities are sufficient to create localized states via the sum rule
(\ref{cond}), under Dirichlet conditions, localization 
seems to be linked to the occurence of confined regions in the
system.

\unitlength 1.85mm
\vspace*{0mm}
{
\begin{figure}
\begin{picture}(80,40)
\def\epsfsize#1#2{0.35#1}
\put(1,5){\epsfbox{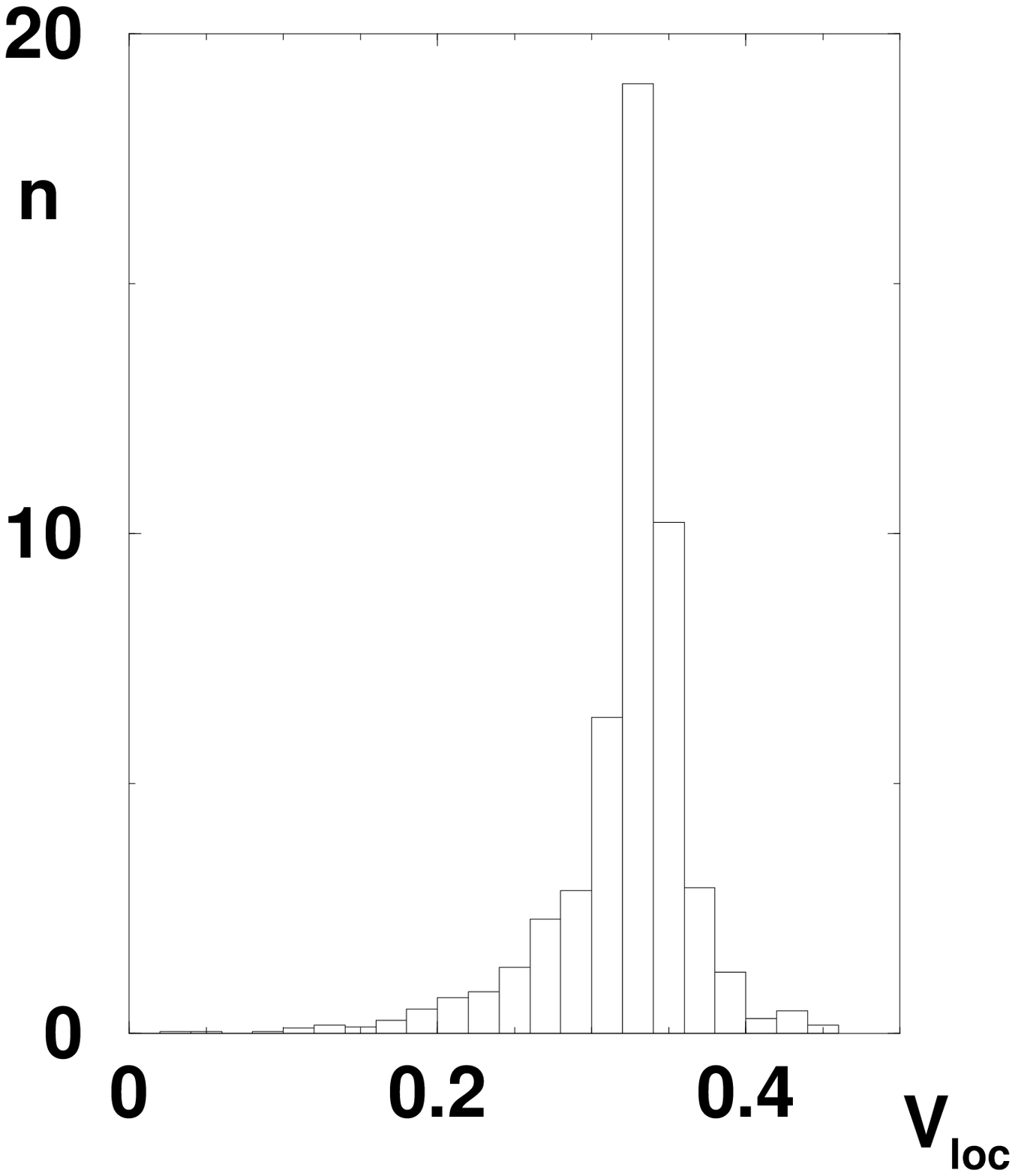}}
\def\epsfsize#1#2{0.4#1}
\put(32,0){\epsfbox{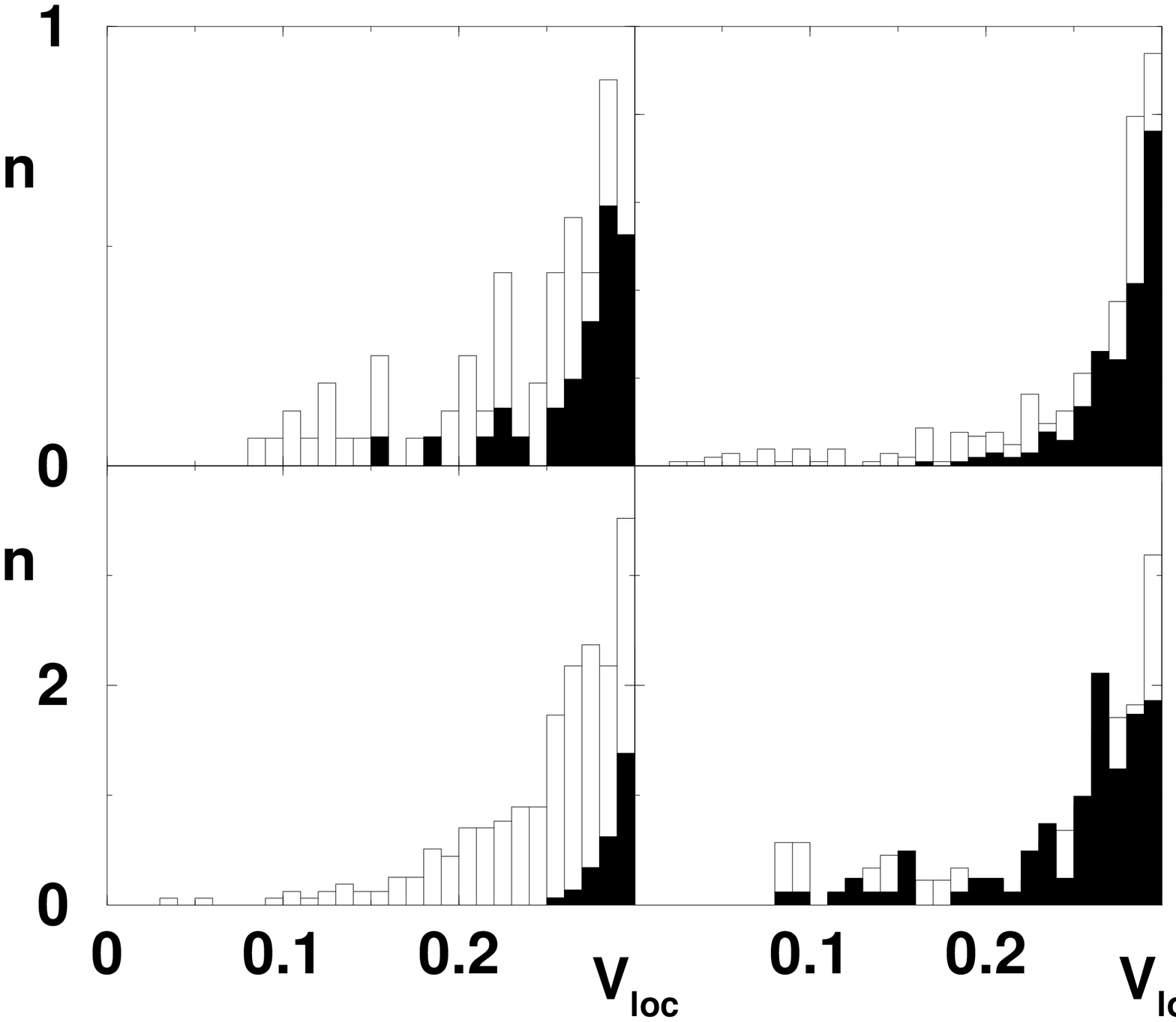}} 
\put(5,31){\makebox(1,1){\bf\Large (a)}}   
\put(34,33){\makebox(1,1){\bf\Large (b)}}  
\put(54,33){\makebox(1,1){\bf\Large (c)}}                                             
\put(34,17){\makebox(1,1){\bf\Large (d)}} 
\put(54,17){\makebox(1,1){\bf\Large (e)}}  
\put(54,23){\unitlength0.1mm
\linethickness{0.4pt} 
\put(0,0){\line(1,0){140}}
\put(0,41){\line(1,0){5}}
\put(5,61){\line(1,0){5}}
\put(10,81){\line(1,0){5}}
\put(15,101){\line(1,0){5}}
\put(20,121){\line(1,0){5}}
\put(5,41){\line(0,1){20}}
\put(10,61){\line(0,1){20}}
\put(15,81){\line(0,1){20}}
\put(20,101){\line(0,1){20}}
\put(25,121){\line(0,1){20}}
\put(25,141){\line(1,0){25}}
\put(50,121){\line(0,1){20}}
\put(65,101){\line(0,1){20}}
\put(80,81){\line(0,1){20}}
\put(95,61){\line(0,1){20}}
\put(110,41){\line(0,1){20}}
\put(125,21){\line(0,1){20}}
\put(125,21){\line(1,0){15}}
\put(110,41){\line(1,0){15}}
\put(95,61){\line(1,0){15}}
\put(80,81){\line(1,0){15}}
\put(65,101){\line(1,0){15}}
\put(50,121){\line(1,0){15}}
\put(0,0){\line(0,1){40}}
\put(140,0){\line(0,1){21}}
}
\put(33,26){\unitlength0.13mm
\linethickness{0.4pt} 
\put(0,0){\line(1,0){110}}
\put(0,41){\line(1,0){30}}
\put(30,79){\line(1,0){80}}
\put(0,0){\line(0,1){41}}
\put(110,0){\line(0,1){79}}
\put(30,41){\line(0,1){38}}
}  
\put(34,7){\unitlength0.07mm
\linethickness{0.4pt} 
\put(0,0){\line(1,0){253}}
\put(0,0){\line(0,1){220}}
\put(253,0){\line(0,1){28}}
\put(0,220){\line(1,0){35}}
\put(35,228){\line(1,0){18}}
\put(53,220){\line(1,0){12}}
\put(65,228){\line(1,0){12}}
\put(77,220){\line(1,0){14}}
\put(91,228){\line(1,0){13}}
\put(104,220){\line(1,0){10}}
\put(114,228){\line(1,0){18}}
\put(132,220){\line(1,0){15}}
\put(147,228){\line(1,0){13}}
\put(160,220){\line(1,0){21}}
\put(181,228){\line(1,0){18}}
\put(199,220){\line(1,0){10}}
\put(209,228){\line(1,0){14}}
\put(223,220){\line(1,0){14}}
\put(237,228){\line(1,0){8}}
\put(35,220){\line(0,1){8}}
\put(53,220){\line(0,1){8}}
\put(65,220){\line(0,1){8}}
\put(77,220){\line(0,1){8}}
\put(91,220){\line(0,1){8}}
\put(104,220){\line(0,1){8}}
\put(114,220){\line(0,1){8}}
\put(132,220){\line(0,1){8}}
\put(147,220){\line(0,1){8}}
\put(160,220){\line(0,1){8}}
\put(181,220){\line(0,1){8}}
\put(199,220){\line(0,1){8}}
\put(209,220){\line(0,1){8}}
\put(223,220){\line(0,1){8}}
\put(237,220){\line(0,1){8}}
\put(245,220){\line(0,1){8}}
\put(245,28){\line(0,1){18}}
\put(253,46){\line(0,1){14}}
\put(245,60){\line(0,1){12}}
\put(253,72){\line(0,1){20}}
\put(245,92){\line(0,1){14}}
\put(253,106){\line(0,1){16}}
\put(245,122){\line(0,1){14}}
\put(253,137){\line(0,1){15}}
\put(245,152){\line(0,1){13}}
\put(253,165){\line(0,1){14}}
\put(245,179){\line(0,1){17}}
\put(253,195){\line(0,1){16}}
\put(245,211){\line(0,1){16}}
\put(245,28){\line(1,0){8}}
\put(245,46){\line(1,0){8}}
\put(245,60){\line(1,0){8}}
\put(245,72){\line(1,0){8}}
\put(245,92){\line(1,0){8}}
\put(245,106){\line(1,0){8}}
\put(245,122){\line(1,0){8}}
\put(245,137){\line(1,0){8}}
\put(245,152){\line(1,0){8}}
\put(245,165){\line(1,0){8}}
\put(245,179){\line(1,0){8}}
\put(245,195){\line(1,0){8}}
\put(245,211){\line(1,0){8}}
} 
\put(5,20){\unitlength0.07mm
\linethickness{0.4pt} 
\put(0,0){\line(1,0){253}}
\put(0,0){\line(0,1){220}}
\put(253,0){\line(0,1){28}}
\put(0,220){\line(1,0){35}}
\put(35,228){\line(1,0){18}}
\put(53,220){\line(1,0){12}}
\put(65,228){\line(1,0){12}}
\put(77,220){\line(1,0){14}}
\put(91,228){\line(1,0){13}}
\put(104,220){\line(1,0){10}}
\put(114,228){\line(1,0){18}}
\put(132,220){\line(1,0){15}}
\put(147,228){\line(1,0){13}}
\put(160,220){\line(1,0){21}}
\put(181,228){\line(1,0){18}}
\put(199,220){\line(1,0){10}}
\put(209,228){\line(1,0){14}}
\put(223,220){\line(1,0){14}}
\put(237,228){\line(1,0){8}}
\put(35,220){\line(0,1){8}}
\put(53,220){\line(0,1){8}}
\put(65,220){\line(0,1){8}}
\put(77,220){\line(0,1){8}}
\put(91,220){\line(0,1){8}}
\put(104,220){\line(0,1){8}}
\put(114,220){\line(0,1){8}}
\put(132,220){\line(0,1){8}}
\put(147,220){\line(0,1){8}}
\put(160,220){\line(0,1){8}}
\put(181,220){\line(0,1){8}}
\put(199,220){\line(0,1){8}}
\put(209,220){\line(0,1){8}}
\put(223,220){\line(0,1){8}}
\put(237,220){\line(0,1){8}}
\put(245,220){\line(0,1){8}}
\put(245,28){\line(0,1){18}}
\put(253,46){\line(0,1){14}}
\put(245,60){\line(0,1){12}}
\put(253,72){\line(0,1){20}}
\put(245,92){\line(0,1){14}}
\put(253,106){\line(0,1){16}}
\put(245,122){\line(0,1){14}}
\put(253,137){\line(0,1){15}}
\put(245,152){\line(0,1){13}}
\put(253,165){\line(0,1){14}}
\put(245,179){\line(0,1){17}}
\put(253,195){\line(0,1){16}}
\put(245,211){\line(0,1){16}}
\put(245,28){\line(1,0){8}}
\put(245,46){\line(1,0){8}}
\put(245,60){\line(1,0){8}}
\put(245,72){\line(1,0){8}}
\put(245,92){\line(1,0){8}}
\put(245,106){\line(1,0){8}}
\put(245,122){\line(1,0){8}}
\put(245,137){\line(1,0){8}}
\put(245,152){\line(1,0){8}}
\put(245,165){\line(1,0){8}}
\put(245,179){\line(1,0){8}}
\put(245,195){\line(1,0){8}}
\put(245,211){\line(1,0){8}}
}  
 
\put(53,9){\unitlength0.15mm
\linethickness{0.4pt} 
\put(0,0){\line(1,0){110}}
\put(0,30){\line(1,0){50}}
\put(50,79){\line(1,0){60}}
\put(0,0){\line(0,1){31}}
\put(110,0){\line(0,1){79}}
\put(50,30){\line(0,1){49}}
}                                                                                                                                                                           
\end{picture}
\caption[]{\small 
(a) Complete histogram of the tooth system 
under Neumann boundary conditions in the frequency regime $0<\omega^2<0.4k/m$.
One can recognize the maximum of the histogram close to the modes with Gaussian
distributions, i.e. with $V_{\rm{loc}}\approx 0.33$.
(b-e) Tails of the histograms of $V_{\rm{loc}}$ in the range 
between $V_{\rm{loc}}=0$ and $0.3$ for (b) the L-shaped system,
(c) the triangular system, (d) the tooth system and
(e) the confined L-shaped system with a much longer and narrower
$A$ region. White columns refer to Neumann, black columns to Dirichlet 
boundary conditions. 
}
\label{bi:hists}
\end{figure}}

\section{Summary and Conclusions}

In summary, we have investigated systems with non-fractal rough
boundaries and computed the relative localization volumes $V_{\rm{loc}}$ of the 
eigenstates.
We found that under Neumann boundary conditions, many localized states are 
coexisting with extended states in the low-frequency regime, whereas under 
Dirichlet conditions localized states are very rare. This difference can be
explained by a sum rule that only applies under Neumann boundary conditions and
implies a localization mechanism for systems with rough boundaries.
By investigating systems of different shapes, we found that number and relative 
localization volumes of the localized states can be triggered by the system geometry, 
i.e. by the complexity of the boundaries and the size ratio of the body 
and the substructures.

The above arguments do not apply, when the 
side lengths of the boundary rectangles are integer multiples of the
main body and the mountains and valleys of the boundary modes fit also into the body 
and therefore cancel under the integral (\ref{cond}), even 
without any decay.
However, in natural systems, where the boundary structures are no simple 
rectangles but random structures, this case will hardly occur.

The sum rule (\ref{cond}) is easier to fulfill in the high-frequency
range, where the mountains and valleys are small and 
numerous. Accordingly, the described localization mechanism 
only applies below some limiting frequency regime whereas
in the high-frequency regime, the localization volumes
are close to the Gaussian value of $V_{\rm{loc}}\approx 0.33$.
There is no crossover between the regimes, but rather a disappearance
of the localized ones around
frequencies between $1.5 k/m \le \omega^2 \le 2k/m$, which refer to a
half-wavelength $\lambda/2\approx 2-3$ lattice constants $a$, the exact value 
depending on details of the
system. This dependence on $a$ in discretized systems is straightforward, as $a$ 
is the smallest common divisor of the different boundary substructures and the
main body of the system. 
In continuous systems, we expect that the localization mechanism also terminates
at wavelengths comparable to the value of the smallest common divisor of
the different lengths of the boundary roughness.

\section{Acknowledgements} We would like to thank the Deutsche 
Forschungsgemeinschaft for financial support.


\begin{thebibliography} {50}
\bibitem{footnote} We did not find any eigenmodes of rectangle $A$ with $n>0$, 
which
is due to the lack of a boundary condition at the right side of the rectangle 
(dashed line in Fig.~\ref{bi:2step}(a)).
\end{thebibliography}
\end{document}